# A HIGGS-FREE MODEL FOR FUNDAMENTAL INTERACTIONS[*]
## PART I: Formulation of the Model

**Marek Pawłowski**[1][†]

Soltan Institute for Nuclear Studies, Warsaw, POLAND

and

**Ryszard Rączka**[2][‡]

Soltan Institute for Nuclear Studies, Warsaw, POLAND

and

Interdisciplinary Laboratory for Natural and Humanistic Sciences
International School for Advanced Studies (SISSA), Trieste, ITALY

May, 1995

## Abstract

We show that the local conformal group $C$ is a natural symmetry group of strong, electroweak and gravitational interactions. A model for these interactions invariant under the local symmetry group $G = SU(3) \times SU(2)_L \times U(1) \times C$ is postulated. It contains all Standard Model and gravitational fields, however the Higgs mass term $\mu^2 \Phi^\dagger \Phi$ is forbidden. Using the unitary gauge and the conformal scale fixing conditions we can eliminate all four real components of the Higgs field in this model. In spite of that the tree level masses of vector mesons, leptons and quarks are automatically generated and they are given by the same formulas as in the conventional Standard Model. The gravitational sector of the model is also analyzed and it is shown that the model admits in the classical limit the Einsteinian form of gravitational interactions. We propose several experimental tests which can discriminate between the Higgs-Free and the Standard Models.

---

[*] This paper is the refined version of the preprint ILAS/EP-1-1995 (hep-ph 9501370), which takes into account the newest electroweak data and is now published in two parts: Part I: Formulation of the model, Part II: Predictions for electroweak observables.

[1] Partially supported by Grant No. 2 P302 189 07 of Polish Committee for Scientific Researches.

[2] Partially supported by the Stiftung Für Deutsch-Polnische Zusammenarbeit Grant No. 984/94/LN.

[†] e–mail: PAWLOWSK@fuw.edu.pl

[‡] e–mail: RRACZKA@fuw.edu.pl

# 1 Introduction

The conventional formulation of Standard Model (SM) is based on two main assumptions:
$1^0$ Gauge symmetry $SU(3) \times SU(2)_L \times U(1)$
$2^0$ The existence of Higgs field and the associated mass generation of vector mesons, leptons and quarks.

The electroweak gauge symmetry was confirmed by experiment by finding the $W_\pm$ and $Z_0$ vector mesons with the predicted masses. However the Higgs particle was not found directly or indirectly in spite of intensive search in tens of experiments [1, 2]. Hence one might suspect that the physical Higgs particle does not exist. There might be a very natural reason for that. In fact the original complex Higgs doublet has four real fields. The three real Higgs fields are eliminated using the $SU(2)_L$ gauge symmetry. If there would be an additional local symmetry then the last surviving Higgs field could be also eliminated.

We show in this work that such a scenario may be indeed realized under the condition that one joints to strong and electroweak interactions also the gravitational interactions. This extension of the class of SM interactions is very natural; indeed the masses of particles are in fact the gravitational charges which produce the gravitational field $g_{\alpha\beta}$. Hence, whenever we have the strong and electroweak interactions of elementary particles, nuclea, atoms or other objects we have also at the same time the gravitational interactions. It seems natural therefore to consider a unified model for strong, electroweak and gravitational interactions which would describe simultaneously all four fundamental interactions. It is well known that gravitational interactions give a negligible effect to most of strong or electroweak elementary particle processes. We show however that they may play the crucial role in a determination of the set of the physical fields and their masses in the unified model and that their presence allows to eliminate all Higgs fields from the final lagrangian.

In order to construct a unique form of the theory of strong and electroweak interactions extended by the gravitational interactions we observe that the original gauge symmetry $SU(3) \times SU(2)_L \times U(1)$ of the fundamental interactions may be naturally extended by the local conformal symmetry. The choice of the unitary gauge condition for $SU(2)_L$ gauge group allows to eliminate the three out of four real Higgs fields from the complex Higgs doublet. In turn the choice of the scale fixing condition connected with the local conformal symmetry allows to eliminate the last Higgs field. In that manner all four Higgs fields can be gauged away completely! It is remarkable that in spite of the elimination of all Higgs fields in our model the vector meson, lepton and quark masses are generated and at the tree level they are given by the same analytical formulas as in the conventional SM.

Thus it may be that the dynamical real Higgs field and the associated Higgs particles are in fact absent and it is therefore not surprising that they could not be detected in various experiments [1, 2].

In Section 2 we discuss the properties of local conformal symmetry and its representations in field space of arbitrary spin. We present in Section 3 the form of the total lagrangian of our unified theory of electroweak, strong and gravitational interactions determined by the gauge and the local conformal invariance. The noteworthy feature of the obtained lagrangian is the lack of the Higgs mass term $\mu^2 \Phi^\dagger \Phi$. We show next that using the unitary gauge condition and the conformal scale fixing condition we can eliminate all dynamical Higgs fields from the theory! We show in Section 4 that in spite of the lack of dynamical Higgs fields the masses of vector mesons, leptons and quarks are generated and at the tree level they are given by the same analytical expressions in terms of coupling constants as in the conventional SM. We give in this section the path integral formulation of our model and show a remarkable result that conformal invariant functions of field operators have the conformal invariant vacuum expectation values.

We present in Section 5 the analysis of the gravitational sector in the unified model. We show that our unified model – after determination of the unitary gauge and scale fixing – leads already on the classical level to the conventional gravitational theory with the Einstein lagrangian implied by the conformal Penrose term contained in the unified lagrangian.

We discuss in Section 6 the possibilities for experimental discrimination between the conventional SM and our HFM. First we show that there are planned some experiments – like the studies of Z-meson pair production in gluon-gluon scattering [3] – for which SM and HFM give essentially different predictions for the cross section considered as the function of 2Z-invariant mass $M_{ZZ}$. Next we call attention to even more fundamental fact that the analysis of electroweak processes at various energy scales may lead to the internal contradictions within the frame-work of SM which may have a natural explanation in the frame-work of HFM. Namely we present in Section 6 the arguments that the value of "Higgs mass" predicted from the analysis of different experimental data is not process independent and that it increases if energy of the considered process increases. We show the collection of experimental data which seems to support our hypothesis. If this observation will be confirmed by future detailed experimental and theoretical analysis this will support our conjecture that the Higgs particle does not exist but the parameter $m_H$ plays the role of UV cutoff.

Finally we discuss in Section 7 several basic problems connected with a description of fundamental interactions which are given by our nonrenormalizable Higgs-free model. We enumerate also some open problems which should be investigated in the framework of HFM.

The present work is the extension of our three previous papers [4],[5],[6] and contains the answer to several questions raised by theirs readers. The predictions of the HFM for almost all electroweak observables are given in Part II of our work [7]. The predictions of our model for the planned LHC experiments and their comparison with the Standard Model predictions will be presented in Part III of our work [8].

## 2 Local conformal symmetry

Let $M^{3,1}$ be the pseudo–Riemannian space time with the metric $g_{\alpha\beta}$ with the signature $(+,-,-,-)$. Let $\Omega(x)$ be a strictly positive function on $M^{3,1}$ which has the inverse $\Omega^{-1}(x)$. Then the local conformal transformation in $M^{3,1}$ is defined as the transformation which changes the metric by the formula

$$g_{\mu\nu}(x) \to \tilde{g}_{\mu\nu}(x) = \Omega^2(x)g_{\mu\nu}(x). \tag{2.1}$$

The set of all local conformal transformations forms the multiplicative abelian infinite–dimensional group $C$ with the obvious group multiplication law.

It should be stressed that a conformal transformation is not a diffeomorphism of the space time. The physical meaning of the conformal transformations follows from the transformation law of the length element

$$dl(x) = \sqrt{-g_{ij}dx^i dx^j} \quad \to \quad \tilde{dl}(x) = \Omega(x)dl(x). \tag{2.2}$$

Hence a local conformal transformation changes locally the length scale. Since the physical phenomena should be independent of the unit chosen locally for the length, time, mass etc. the group $C$ of local conformal transformations should be a symmetry group of physical laws.

We shall give now a construction of the representation of the conformal group $C$ in the field space. Let $\Psi$ be a tensor or spinor field of arbitrary spin. Define the map

$$\Omega \to U(\Omega)$$

by the formula

$$\tilde{\Psi}(x) = U(\Omega)\Psi(x) = \Omega^s(x)\Psi(x), \qquad s \in R \tag{2.3}$$

The number $s$ is the conformal weight of $\Psi$ determined by the condition of conformal invariance of field equation. It is evident that the map $\Omega \to U(\Omega)$ defines the representation of $C$ in the field space.

Using the above definitions one can show that the Maxwell $F_{\mu\nu}$ and the Yang–Mills $F_{\mu\nu}{}^a$ field strengths on $(M^{3,1}, g)$ have the conformal weight $s = 0$ whereas the massless Dirac field has the conformal weight $s = -\frac{3}{2}$ [9]-[11]. It is noteworthy that the scalar massless field $\Phi$ satisfying the Laplace–Beltrami equation

$$\triangle \Phi = 0$$

is not conformal invariant. In fact it was discovered by Penrose that one has to add to the Lagrangian on $(M^{3,1}, g)$ the term

$$-\frac{1}{6} R \Phi^\dagger \Phi$$

where $R$ is the Ricci scalar, in order that the corresponding field equation is conformal invariant with the conformal weight $s = -1$ [12].

## 3 A unified model for strong, electroweak and gravitational interactions

We postulate that the searched unified theory of strong, electroweak and gravitational interactions will be determined by the condition of invariance with respect to the group $G$

$$G = SU(3) \times SU(2)_L \times U(1) \times C \tag{3.1}$$

where $C$ is the local conformal group defined by (2.1). Let $\Psi$ be the collection of vector meson, fermion and scalar fields which appear in the conventional minimal SM for electroweak and strong interactions. Then the minimal natural conformal and $SU(3) \times SU(2)_L \times U(1)$ –gauge invariant total lagrangian $L(\Psi)$ may be postulated in the form:

$$L = [L_G + L_F + L_Y + L_\Phi + L_{grav}]\sqrt{-g} \tag{3.2}$$

Here $L_G$ is the total lagrangian for the gauge fields $A_\mu^a$, $W_\mu^b$ and $B_\mu$, $a = 1, ..., 8$, $b = 1, 2, 3$ associated with $SU(3) \times SU(2)_L \times U(1)$ gauge group

$$L_G = -\frac{1}{4} F^a{}_{\mu\nu} F^{a\,\mu\nu} - \frac{1}{4} W^b{}_{\mu\nu} W^{b\,\mu\nu} - \frac{1}{4} B_{\mu\nu} B^{\mu\nu}, \tag{3.3}$$

and $F^a{}_{\mu\nu}$, $W^b{}_{\mu\nu}$ and $B_{\mu\nu}$ are the conventional field strengths of gauge fields in which the ordinary derivatives are replaced by the covariant derivatives e.g.

$$B_{\mu\nu} = \nabla_\mu B_\nu - \nabla_\nu B_\mu, \tag{3.4}$$

etc.; $L_F$ is the lagrangian for fermion field interacting with the gauge fields; $L_Y$ represents the Yukawa interactions of fermion and scalar fields; $L_\Phi$ is the $G$-invariant lagrangian for the scalar fields, which may be written in the form:

$$L_\Phi = (D\Phi)^\dagger (D\Phi) - \lambda(\Phi^\dagger \Phi)^2 + \beta \partial_\mu |\Phi| \partial^\mu |\Phi| - \frac{1}{6}(1+\beta) R \Phi^\dagger \Phi, \qquad (3.5)$$

where $D$ denotes the $SU(2)_L \times U(1)$ covariant derivative which is covariant also with respect to the dynamical curved space and $\beta$ is an arbitrary real number, in general. Notice that the condition of conformal invariance does not admit the Higgs mass term $\mu^2 \Phi^\dagger \Phi$ which assures the mechanism of spontaneous symmetry breaking and mass generation in the conventional formulation. Instead we have two additional terms: the Penrose term

$$-\frac{1}{6}(1+\beta) R \Phi^\dagger \Phi \qquad (3.6)$$

which assures that the lagrangian (3.5) is conformal invariant, and the term

$$\beta \partial_\mu |\Phi| \partial^\mu |\Phi| \qquad (3.7)$$

which together with the term $-\frac{1}{6}\beta R \Phi^\dagger \Phi$ is conformal and gauge invariant. It may be surprising that (3.7) depends on $|\Phi|$. Observe however that the conventional first term in $L_\Phi$ can be written in the form

$$(D\Phi)^\dagger (D\Phi) = \partial_\mu |\Phi| \partial^\mu |\Phi| + |\Phi|^2 L_\sigma(g(\Phi), W, B) \qquad (3.8)$$

where $g(\Phi)$ is $SU(2)_L$ gauge unitary matrix defined by the formula

$$\Phi = \begin{pmatrix} \phi_u \\ \phi_d \end{pmatrix} = g(\Phi) \begin{pmatrix} 0 \\ |\Phi| \end{pmatrix}, \quad g(\Phi) = \frac{1}{|\Phi|} \begin{pmatrix} \bar{\phi}_d & \phi_u \\ -\bar{\phi}_u & \phi_d \end{pmatrix} \qquad (3.9)$$

and $L_\sigma(g(\Phi), W, B)$ is a gauged–sigma–model–like lagrangian.

We see therefore that the term like (3.7) is already present in the conventional gauge invariant lagrangian.

The last term in (3.2) is the Weyl term

$$L_{grav} = -\rho C^2, \qquad \rho \geq 0, \qquad (3.10)$$

where $C^\delta_{\alpha\beta\gamma}$ is the Weyl tensor which is conformally invariant. Using the Gauss–Bonnet identity we can write $C^2$ in the form

$$C^2 = 2(R^{\mu\nu} R_{\mu\nu} - \frac{1}{3} R^2). \qquad (3.11)$$

We see that the condition of conformal invariance does not admit in (3.2) the conventional gravitational Einstein lagrangian

$$L = \kappa^{-2} R \sqrt{-g}, \qquad \kappa^2 = 16\pi G. \tag{3.12}$$

We show however in Section 5 that the Einstein lagrangian is reproduced from (3.6) when the scale fixing condition is chosen properly. Note that the quantum gravity sector contained in (3.2) is perturbatively renormalizable [13] whereas the quantum gravity defined by the Einstein lagrangian (3.12) coupled with matter is nonrenormalizable [14]. Hence, for a time being it is an open question which form of gravitational interaction is more proper on the quantum level. The discussion of the role of quantum effects which may reproduce the lagrangian (3.12) and give the classical Einstein theory as the effective induced gravity was presented in our previous work [5].

Notice that conformal symmetry implies that all coupling constants in the present model are dimensionless.

The theory given by (3.2) is our conformally invariant proposition alternative to the standard Higgs–like theory with SSB. Its new, most important feature is the local conformal invariance. It means that simultaneous rescaling of all fields (including the field of metric tensor) with a common, arbitrary, space–time dependent factor $\Omega(x)$ taken with a proper power for each field (the conformal weight) will leave the Lagrangian (3.2) unaffected. The symmetry has a clear and obvious physical meaning [15], [10]. It changes in every point of the space–time all dimensional quantities (lengths, masses, energy levels, etc) leaving theirs ratios unchanged. It reflexes the deep truth of the nature that nothing except the numbers has an independent physical meaning.

The freedom of choice of the length scale is nothing but the scale fixing freedom connected with the conformal symmetry group. In the conventional approach we define the length scale in such a way that elementary particle masses are the same for all times and in all places. This will be the case when we rescale all fields with the $x$–dependent conformal factor $\Omega(x)$ in such a manner that the length of the rescaled scalar field doublet is fixed i.e.

$$\tilde{\Phi}^\dagger \tilde{\Phi} = \frac{v^2}{2} = const. \tag{3.13}$$

Let us note that if we would introduce instead of $C$-symmetry the additional local symmetry in the form of a unitary or rotation gauge group then its unitary representation in the field space could change only phase but not the modulus of the Higgs doublet. We see therefore that the local conformal group which changes the modulus of Higgs field $\Phi$ is ideally suited for the elimination of all Higgs fields from the theory of electroweak and strong interactions.

Obviously we can choose other than (3.13) scale fixing condition, e.g. we can use the freedom of conformal factor to set

$$\sqrt{-\tilde{g}} = 1; \tag{3.14}$$

this will lead to other local scales but as we show below it will leave physical predictions unchanged.

It follows from Fadeev-Popov method that the expectation values of gauge invariant function of field operators are gauge invariant i.e. they are independent from a chosen gauge fixing condition. We shall derive now the analogous result for the local conformal group and show that the expectation values of conformal invariant operators are independent on the choice of scale fixing condition.

In order to show this we shall use the functional integral formalism. Let $L[\Psi]$ be the scale invariant lagrangian (3.2). Let $C(\Psi)$ be the function of field operators which is local conformal invariant i.e.

$$C(\tilde{\Psi}) = C(\Psi)$$

where $\tilde{\Psi} = \Omega^{s_\Psi}\Psi$ is the conformal transform of scalar, vector or fermion field respectively given by (2.3) and determined by their conformal degree $s_\Psi$. Then according to the so called Matthews theorem the path integral representation for vacuum expectation values of $C(\Psi)$ has the form [16]:

$$<C(\Psi)>_0 = Z^{-1}\int C(\Psi)e^{iS_T(\Psi)}\Delta_f(\Psi)\delta[f(\Psi)]D\Psi \tag{3.15}$$

where $Z$ is the partition function

$$Z = \int e^{iS_T(\Psi)}\Delta_f(\Psi)\delta[f(\Psi)]D\Psi \tag{3.16}$$

$S_T = S + S_{FP}$ where $S_{FP}$ is the Fadeev-Popov contribution to the action integral due to the gauge fixing conditions and $f(\Psi)$ is the scale fixing condition. $D\Psi$ is the functional measure over all dynamical fields in $\Psi$ and in our case has the form

$$D\Psi = D\Phi DAD\psi Dg \tag{3.17}$$

We chose the gauge fixing condition in such a manner that $S_{FP}$ is conformal invariant. It follows from Fadeev-Popov formalism [17] that

$$\Delta_f(\Psi)\int \delta[f(\Psi^\Omega)]D\Omega = I \tag{3.18}$$

where $D\Omega$ is the invariant measure on the conformal group and is given by the formula
$$D\Omega = \prod_x \frac{d\Omega(x)}{\Omega(x)} \qquad (3.19)$$
One readily verifies that this measure is invariant under the group multiplication $\Omega \to \Omega'\Omega$ and the inversion $\Omega \to \Omega^{-1}$.

It follows from the conformal invariance of $D\Omega$ that $\Delta_f(\Psi)$ is conformal invariant. Setting as in (3.13)
$$f(\Psi^\Omega) = |\Phi^\Omega| - \frac{v}{\sqrt{2}}$$
and using the measure invariance we obtain
$$\Delta_f(\Psi) = \frac{v}{\sqrt{2}}$$

We present now the important result:
**Theorem 4.1**
Let $C(\Psi)$ be the conformal invariant function of field operators. Then the vacuum expectation value $< C(\Psi) >_0$ given by (3.15) is independent on the scale fixing condition.

(For the proof see Appendix.)

This result is a little bit surprising, especially if one takes into account how different are the scale fixing conditions (3.13) and (3.14). Theorem 4.1 implies that we can calculate the vacuum expectation values of conformal invariant function of field operators using the most convenient scale fixing condition. Since the condition (3.13) together with the unitary gauge fixing condition for $SU(2)_L$ group eliminates all four Higgs fields from the action integral $S_T(\Psi)$ we shall use it exclusively in all following calculations. We note that the scattering operator $\hat{S}$ is dimensionless and therefore conformal invariant. Consequently if we use the normalization of asymptotic states such that they are dimensionless we can use the scale fixing condition (3.13) for calculation of probability amplitudes of all physical processes.

# 4 Generation of lepton, quark and vector boson masses

We demonstrate now that using the conformal group scale fixing condition (3.13) we can generate the same lepton, quark and vector meson masses as in the conventional SM without however use of any kind of Higgs mechanism and SSB.

In fact inserting the scale fixing condition (3.13) into the Lagrangian (3.2) we obtain

$$\tilde{L} = L^{scaled} = [L_G + L_F + L_Y^{scaled} + L_\Phi^{scaled} + L_{grav}]\sqrt{-g}, \qquad (4.1)$$

in which the condition (3.13) was inserted into $L_\Phi$ and $L_Y$. We should use the symbol $\tilde{\Phi}$, $\tilde{\Psi}$ etc. for the rescaled fields in (4.1), however for the sake of simplicity we shall omit " ˜ " sign over fields in the following considerations.

The condition (3.13) together with the unitary gauge fixing of $SU(2)_L \times U(1)$ gauge group, reduce by (3.9) the Higgs doublet to the form

$$\Phi^{scaled} = \frac{1}{\sqrt{2}}\begin{pmatrix} 0 \\ v \end{pmatrix}, \qquad v > 0 \qquad (4.2)$$

and produce the tree level mass terms for leptons, quarks and vector bosons associated with $SU(2)_L$ gauge group. For instance the $\Phi$–lepton Yukawa interaction $L_Y^l$ reads

$$L_Y^l = -\sum_{i=e,\mu,\tau} G_i \bar{l}_{iR}(\Phi^\dagger l_{iL}) + h.c.$$

where

$$l_{eL} = \begin{pmatrix} \nu_e \\ e_L \end{pmatrix} \quad etc.$$

It passes into

$$L_Y^{l\ scaled} = -\frac{1}{\sqrt{2}}v(G_e \bar{e}e + G_\mu \bar{\mu}\mu + G_\tau \bar{\tau}\tau) \qquad (4.3)$$

giving the conventional, space–time independent lepton masses

$$m_e = \frac{1}{\sqrt{2}}G_e v, \qquad m_\mu = \frac{1}{\sqrt{2}}G_\mu v, \qquad m_\tau = \frac{1}{\sqrt{2}}G_\tau v. \qquad (4.4)$$

Similarly one generates from $\Phi$–quark Yukawa interaction $L_Y^q$ the corresponding quark masses. In turn from $L_\Phi$-lagrangian (3.5) using the scaled scalar field (4.2) one obtains

$$(D_\mu \Phi)^\dagger D^\mu \Phi = \frac{g_2^2 v^2}{4}W_\mu^+ W^{\mu-} + \frac{g_1^2 + g_2^2}{8}v^2 Z^2$$

where

$$Z_\mu = -\sin\theta_W B_\mu + \cos\theta_W W^3{}_\mu, \qquad \cos\theta_W = \frac{g_2}{\sqrt{g_1^2 + g_2^2}}.$$

Hence one obtains the following vector mesons masses

$$m_W = \frac{v}{2}g_2, \qquad m_Z = \frac{m_W}{\cos\theta_W}. \qquad (4.5)$$

It is remarkable that the analytical form for tree level fermion and vector meson masses in terms of coupling constants and the parameter $v$ is the same as in the conventional SM. We see therefore that the Higgs mechanism and SSB is not indispensable for the fermion and vector mesons mass generation!

We note that the fermion–vector boson interactions in our model are the same as in SM. Hence analogously as in the case of conventional formulation of SM one can deduce the tree level relation between $v$ and $G_F$ – the four–fermion coupling constant of $\beta$–decay:

$$v^2 = (2G_F)^{-1} \to v = 246 GeV. \tag{4.6}$$

Here we have used the standard decomposition $g^{\mu\nu}\sqrt{-g} = \eta^{\mu\nu} + \sqrt{2}\kappa h^{\mu\nu}$ (see e.g. [18]) which reduces the tree level problem for the matter fields to the ordinary flat case task.

We see therefore that the resulting expressions for masses of physical particles are identical as in the conventional SM.

Let us stress that the scale fixing condition like (3.13) does not break $SU(2)_L \times U(1)$ gauge symmetry. The symmetry is broken (or rather one of gauge equivalent description is fixed) when (3.13) is combined with unitary gauge condition of electroweak group leading to (4.2). However, also after imposing of a gauge condition like (4.2) we have a remnant of both the conformal and $SU(3) \times SU(2)_L \times U(1)$ initial gauge symmetries: this is reflected in the special, unique relations between couplings and masses in our model.

Let us note that taking different scale and gauge fixing conditions the number of the dynamical fields in a given sector changes but the total number of independent fields is the same. In fact if we would choose an another scale fixing condition as e.g. given by (3.14) and say the covariant gauge condition then we would have in the gravitational, $SU(2)_L$-gauge and Higgs sector nineteen independent degrees of freedom namely nine $g_{\alpha\beta}$ fields, six $W_\perp^a$ fields and four Higgs fields. If instead we take the scale fixing condition (3.13) and then unitary gauge fixing condition we have ten $g_{\alpha\beta}$ fields, nine $W_l^\pm$ and $Z_l$, $l = 1, 2, 3$ fields and zero Higgs fields i.e. again nineteen independent fields.

## 5  Gravity Sector

Let us impose the scale fixing condition (3.13) on the lagrangian (3.2) and collect all gravitational terms. The lagrangian reads:

$$L^{scaled} = [L^{scaled}_{matter} - \frac{1}{12}(1+\beta)v^2 R - 2\rho(R^{\mu\nu}R_{\mu\nu} - \frac{1}{3}R^2) - \frac{\lambda}{4}v^4]\sqrt{-g} \tag{5.1}$$

where we have selected the part $L_{matter}^{scaled}$ (describing the matter interacting with gravity) from the remaining purely gravitational terms.

The variation of (5.1) with respect to the metric $g^{\mu\nu}$ leads to the following classical equation of motion:

$$\rho[-\frac{2}{3}R_{;\mu;\nu} + 2R_{\mu\nu}{}^{;\eta}{}_{;\eta} - \frac{2}{3}g_{\mu\nu}R^{;\eta}{}_{;\eta} -$$

$$4R^{\eta\lambda}R_{\mu\eta\nu\lambda} + \frac{4}{3}RR_{\mu\nu} + g_{\mu\nu}(R^{\eta\lambda}R_{\eta\lambda} - \frac{1}{3}R^2)]+$$

$$\frac{1}{12}(1+\beta)v^2(R_{\mu\nu} - \frac{1}{2}g_{\mu\nu}R) + \frac{\lambda}{8}v^4 g_{\mu\nu} = \frac{1}{2}T_{\mu\nu}. \quad (5.2)$$

In the empty case $T_{\mu\nu} = 0$ this equation is satisfied by all solutions of an empty space Einstein equation with a properly chosen cosmological constant $\Lambda_c$:

$$R_{\mu\nu} - \frac{1}{2}g_{\mu\nu}R + \Lambda_c g_{\mu\nu} = 0. \quad (5.3)$$

In fact (5.3) implies that

$$R_{\mu\nu} \sim g_{\mu\nu} \quad \Rightarrow \quad R_{\mu\nu} = \frac{1}{4}Rg_{\mu\nu} \quad (5.4)$$

and then

$$R_{\mu\nu} = \Lambda_c g_{\mu\nu}. \quad (5.5)$$

Inserting (5.4) into (5.2) we find that the part proportional to $\rho$ vanishes. The remnant can be collected leading to the relation

$$\frac{1}{8}v^2 g_{\mu\nu}(\frac{2}{3}(1+\beta)\Lambda_c - \lambda v^2) = 0 \quad (5.6)$$

where the empty space condition $T_{\mu\nu} = 0$ were used for the right hand side of (5.6).

Equation (5.6) implies

$$\Lambda_c = \frac{3}{2(1+\beta)}\lambda v^2. \quad (5.7)$$

Equation (5.7) relates the undetermined so far coupling constant $\lambda$ with a potentially observable cosmological constant $\Lambda_c$.

Let us go back to the case with the matter. Observe that the term linear in the curvature appears in (5.1) with the coefficient $-\frac{1}{12}(1+\beta)v^2$. If we want to reproduce the correct gravitational sector already at the classical level we have to admit for nonzero $\beta$ coupling. Setting

$$-\frac{1}{12}(1+\beta)v^2 = \kappa^{-2} \tag{5.8}$$

we reproduce the Newtonian coupling in front of curvature $R$ in (5.1). This would mean that $\beta \approx -10^{38}$! Notice however that taking the scale fixing condition (3.13) the term $\beta \partial_\mu |\Phi| \partial^\mu |\Phi|$ vanishes. Hence it looks like that the only role of this term is to generate the proper value of Newton constant in the Einstein tree level lagrangian resulting from the Penrose term. (For further discussion see [5].)

Let us stress that since in our formalism we do not use the SSB mechanism for mass generation the self-interaction term $\lambda(\Phi^\dagger \Phi)^2$ can be set to zero by setting $\lambda = 0$. In this case the obtained in our formalism cosmological constant (5.7) is also zero in agreement with experiments and the conviction of Einstein and many other authors [19]. Consequently the so called "worst fine-tuning problem in the history of physics" [20]-[22] is solved in our model in a very natural manner.

The cosmological constant $\Lambda_c$ given by (5.7) was obtained from the analysis of gravitational interactions in the empty space-time. In reality the matter is always present and modifies the formula for $\Lambda_c$. In this case the most natural definition of the effective cosmological constant was given by Zel'dovich [23] and by Adler [24] by means of the partition function determined by the lagrangian (4.1). (See also the excellent analysis of this problem in [25].) However – up to now – nobody was able to get any quantitative prediction for the value of $\Lambda_c$ in this frame-work.

# 6 Proposals for a discrimination between Standard and Higgs-Free Models.

We shall analyze now the perspectives for an experimental discrimination between the conventional SM and the HFM. We see at present two possibilities:

I. Consider the processes for which the SM and HFM give drastically different predictions. For instance consider the Z-meson pair production process in the gluon-gluon collision which will be investigated in LHC experiment [3]. In HFM this process in one-loop approximation is described by the quark box diagrams

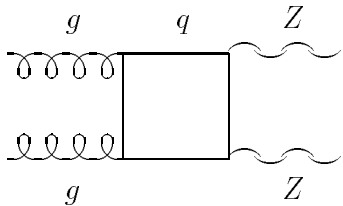

It is noteworthy that in this case the HFM scattering amplitude and the Z-meson pair production cross section can be calculated without any UV cutoff.

This process considered in SM has the additional Higgs exchange Feynman diagrams which give a resonance type contribution [26]

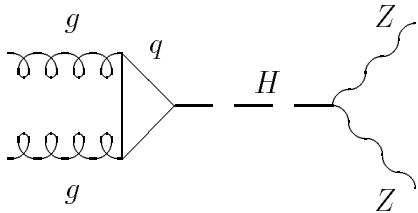

Hence one obtains in SM the characteristic pick in the cross section considered as the function of 2Z invariant mass distribution [26], whereas in HFM one obtains the pickless cross section monotonically decreasing with the invariant mass of the produced Z-boson pair.

The same phenomena will occur e.g. in the case of Z-pair or $W^+W^-$ production in photon-photon high energy scattering [27] or in many other processes considered as candidates for the direct search of the Higgs particle.

The position of the Higgs resonance pick is given by the unknown mass $m_H$ of the searched particle. Assuming that SM is valid we can have some estimation for $m_H$ from the so called precision tests. At present the information from precision tests concerning the Higgs boson mass is still not very conclusive and essentially depends on the assumptions concerning the set of fitted experimental data [28]. We can expect, however that this will change soon mostly due to the increasing accuracy of the top and W mass measurements and the further analysis of LEP1 and SLAC data. The constructed LEP2 and LHC accelerators will be probably able to test directly the region of the Higgs mass predicted by the precision tests. The possible negative result of the direct

Higgs meson searches in the Higgs mass interval admissible by EW precision tests would mean an evident contradiction in the framework of SM. The above contradictions will not appear in HFM since there the Higgs particle is absent. We see therefore that expected in a near future considerable increase of accuracy of EW precision tests and the new accelerator experiments may provide a crucial tests for Standard and Higgs-Free Models.

II. Consider electroweak processes at various energy (or momentum transfer) scales. We wish to show that the results coming from various energy scale experiments may lead to the contradictions within the SM frame-work but they can be naturally explained in the frame-work of HFM.

First we would like to elucidate the role which plays the Higgs mass $m_H$ in SM. It is convenient to use for this purpose the universal parameters $\varepsilon_1$, $\varepsilon_2$, $\varepsilon_3$ introduced by Altarelli *et al.* [29] by means of which the radiative corrections to most of electroweak observables measured in LEP, CDF, D0 or SLC experiments can be expressed. If we calculate these parameters in our model in one loop-approximation we find the specific class of Feynman diagrams with fermion and vector boson loops which contributes to them. Since some vector boson loops will produce divergences, e.g. in the case of fermion – massive vector boson coupling constant, one has to introduce either the new renormalization constants or UV cutoff $\Lambda$ which can be given by the formula [30]

$$\log \frac{\Lambda^2}{\mu^2} = \frac{2}{4-D} - \gamma_E + \log 4\pi + \frac{5}{6} \tag{6.1}$$

where $\mu$ is the reference mass of dimensional regularization, $D$ is the space–time dimension and $\gamma_E$ is the Euler's constant.

One obtains the formula for $\varepsilon_i$ parameters in SM if one adds to the class of Feynman diagrams in our model all appropriate one–loop diagrams with Higgs internal lines. Using the results of [30] and [31] one obtains

$$\varepsilon_1^{SM} - \varepsilon_1^{HFM} = \frac{3\alpha(m_Z^2)}{16\pi c_0^2} \log\left(\frac{\Lambda^2}{m_H^2}\right) + \varepsilon_{1rem}(X)$$

$$\varepsilon_2^{SM} - \varepsilon_2^{HFM} = \varepsilon_{2rem}(X) \tag{6.2}$$

$$\varepsilon_3^{SM} - \varepsilon_3^{HFM} = \frac{\alpha(m_Z^2)}{48\pi s_0^2} \log\left(\frac{\Lambda^2}{m_H^2}\right) + \varepsilon_{3rem}(X)$$

$$X = \frac{m_Z^2}{m_H^2} \log \frac{m_Z^2}{m_H^2}$$

where $HFM$ index of $\varepsilon_i$ means that the quantity was calculated in our Higgs-Free Model. Here $\alpha(m_Z^2) = \frac{1}{128.87}$ and $c_0$ and $s_0$ are defined by the formula

$$s_0^2(1-s_0^2) = s_0^2 c_0^2 \equiv \frac{\pi\alpha(m_Z^2)}{\sqrt{2}G_F m_Z^2}.$$

The functions $\varepsilon_{i\,rem}(X) \to 0$ if $m_H \to \infty$.

The above formulas indicate a role which plays in SM the very large Higgs mass: first the numerical analysis shows that the term $\varepsilon_{i\,rem}(X)$ for $m_H \geq 300 GeV$ can be disregarded and second if we take the UV cutoff $\Lambda \simeq m_H$ then by (6.2) the prediction for $\varepsilon_i$–parameters in the conventional SM and our nonrenormalizable model almost coincide. Thus the very large Higgs mass preferred by the top mass $m_t = 180 GeV$ plays in fact in the conventional SM the role of UV cutoff parameter. If the Higgs particle will be not found then our model provides an extremely natural frame–work for the description of electroweak and strong interactions at least up to TeV energies.

If our interpretation that the Higgs mass $m_H$ is in fact the UV cutoff is correct then the deduced from various experiments "Higgs mass" may vary with energy scale $E$ of the relevant processes. Consequently for instance the value of $m_H$ deduced from parity violation experiment (PVE) [32] for which $\sqrt{-q^2} < 1 MeV$ may be smaller than the value of $m_H$ deduced from LEP, SLAC or Fermilab experiments where $m_Z$ is the characteristic scale of electroweak processes. We illustrate our hypothesis using the effective leptonic weak mixing angle $\theta_W^{eff,l}$.

Averaging over the experimental data [1, 2] we obtain

$$sin^2\theta_W^{eff,l}(\text{LEP + SLAC + Fermilab + UA2}) = 0.2320 \pm 0.0004$$

$$sin^2\theta_W^{eff,l}(\text{PVE}+\nu\text{e}+\nu\text{p + eD}) = 0.225 \pm 0.005$$

The above numbers are not definitively inconsistent however their rather weak consistency allows for consideration of both alternatives: the case where the future, more precise measurements will confirm the consistency of these numbers and the case where the future experiments will detect a divergence between these quantities. The last alternative is interesting for our analysis. If we consider the SM prediction for $sin^2\theta_W^{eff,l}$ in one-loop approximation we obtain that $sin^2\theta_W^{eff,l}$ is a definite function of the Higgs mass $m_H$. This relation can be inverted leading to the relation $m_H(sin^2\theta_W^{eff,l})$ which is plotted in Fig. 6.1.

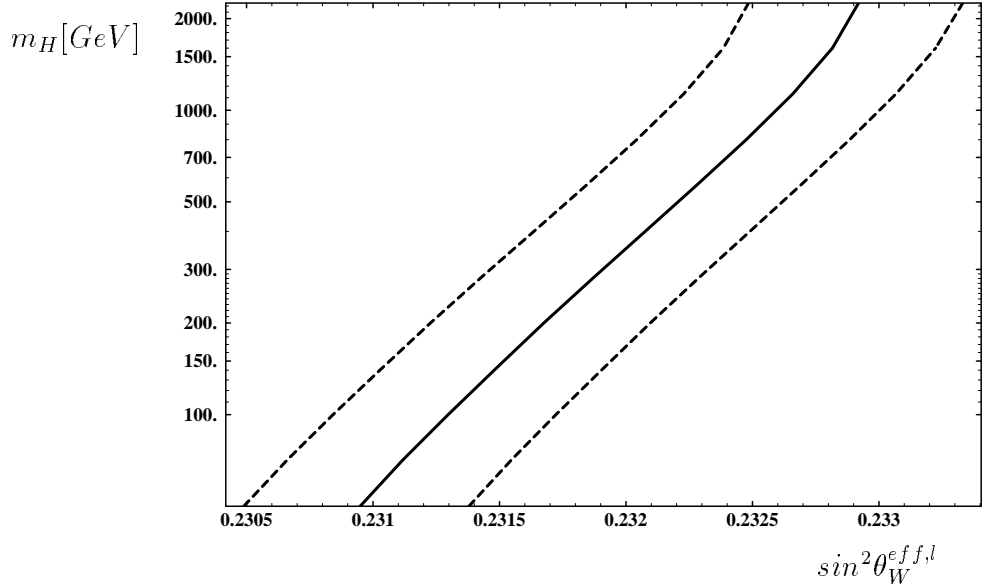

**Fig. 6.1** *The plot of $m_H$ as the function of $sin^2\theta_W^{eff,l}$ for $m_t = 180 GeV$ (solid line) and $m_t = 180 \pm 12 GeV$ (dashed lines).*

We see that $m_H$ increases monotonically with $sin^2\theta_W^{eff,l}$. Hence the possible difference in $sin^2\theta_W^{eff,l}$ predictions would mean that the "Higgs mass" is different for different energy scales.

If the Higgs mass in all formulas for radiative corrections represents the pole mass of real particle then the value $m_H$ obtained from various experiments should be independent of energy scale $E$ of a given experiment. This means that SM predicts

$$m_H(E) = const \tag{6.3}$$

and consequently

$$sin^2\theta_W^{eff,l} = const. \tag{6.4}$$

(The constrain (6.4) holds for most of extensions of SM also.) If our hypothesis will turn out to be true it would mean that the SM predictions (6.3) and (6.4) do not hold, and that the SM is inconsistent. In turn our hypothesis that $m_H$ varies with energy scale $E$ of the process may be easily explained within the HFM. Our model predicts that Higgs particle is absent and $m_H$ just plays a role of UV cutoff $\Lambda$ which should increase if the energy scale increases. This feature of our model is not surprising. We recall that in the well understood theories like QED the value of the UV cutoff $\Lambda$ depends on the energy scale of a given experiment and for low energy phenomena $\Lambda$ can be very small. The famous example is provided by the Bethe treatment of the Lamb's shift for hydrogen atom [33]; Bethe found that the natural relativistic invariant UV cutoff (which implies that the energy integrals for the Lamb's shift are finite) is provided by the electron mass $m_e = 0.51 MeV$, which is still very large with respect to energy scale given by the value $< E_n - E_m >_{av}$. This example illustrates a general principle that for each energy scale one has the energy dependent cutoff $\Lambda$ appropriate to the considered energy.

It was generally assumed up to now in most of reviews and analyses of electroweak experimental data ([34],[1],[2],[29]...) that there is no process and energy dependence of Higgs mass, $\Delta r$, $\Delta \rho$, $\varepsilon_1$, $\varepsilon_2$, $\varepsilon_3$, $\varepsilon_b$ and other universal electroweak parameters. This approach partially followed from the deep believe in Standard Model or its extensions and partially followed from the limited accuracy of individual experiments which does not allow to study this aspect of electroweak interactions. It seems that at least the last obstacle will disappear in the near future. We propose to include to the list of precision tests of electroweak theories the experimental study of energy independence (or dependence) of universal electroweak parameters. We hope that the increasing statistics of Tevatron and SLC data, the expected data from new LEP2 and LHC facilities constructed at CERN and the progress – up to 1% accuracy [35] – in low and medium energy experiments will provide us with sufficient information necessary to study this problem. If such analysis will confirm our hypothesis on energy dependence of the "Higgs mass" then this would exclude SM and most of its extensions and would give a strong evidence that $m_H$ is not a mass of real physical particle but only an energy dependent cutoff like in our HFM.

A similar analysis can be carried out also for energy dependent electroweak observables like asymmetries or cross sections ratios for which one can study the deviations from the SM predicted energy dependence.

## 7 Discussion.

The elementary particle physics is at present at a crossroad. We have in fact three drastically different alternatives:

I$^o$ The Higgs particle exists, its mass will be experimentally determined and it will have the value predicted by the radiative corrections of SM. This will confirm the SSB mechanism for mass generation, the validity of SM framework and it will represent an extraordinary success of quantum gauge field theory.

II$^o$ The Higgs particle exists but its mass is considerable different from that predicted by the radiative corrections of SM. This would signal some kind of "New Physics" which will imply a reformulation of the present version of SM.
III$^o$ The Higgs particle does not exists. This will lead to a rejection of SM with Higgs sector and it will give preference to Higgs-free models for fundamental interactions. Presumably the obtained physical Higgs-free models will be nonrenormalizable.

It may be that the renormalizability of Quantum Gravity determined by Einstein action integral coupled with matter fields is not an "accident at work in quantum field theory" but it represents a universal feature that physical fundamental interactions considered simultaneously are nonrenormalizable. In this situation we are compulsed to use the nonrenormalizable models of quantum field theory for a description of fundamental interactions and we have to learn how to deduce predictions for experiments from such models. Therefore it seems necessary to develop perturbative and nonperturbative methods for extracting predictions for scattering amplitudes and observables from nonrenormalizable massive vector meson models. The theoretical aspects of these problems are discussed in [36] and [37]. The very interesting method of calculation of observables in nonrenormalizable massive vector meson models using the associated effective field theory was developed recently by Herrero and Morales [38] and by Bilenky and Santamaria [39]. One can also use the technique of so called Generalized Equivalence Theorem which was recently extended to nonrenormalizable massive vector meson models [40].

The proposed HFM allows to obtain the Einsteinian form of gravitational interactions in the classical limit. It can be also analyzed by means of effective action for induced gravity [25].

We present in Part II of our work the new method of getting predictions from our nonrenormalizable model [7]. Using it we derived the definite predictions for almost all electroweak observables measured in the recent experiments at LEP1, Tevatron and SLC. The obtained predictions in HFM are in remarkable agreement with experimental data. Consequently the existence of Higgs field and the associated Higgs particle seems not necessary for the explanation of experimental data. Since our model contains less fundamental fields than the conventional SM and is therefore conceptually much simpler we believe that it can be considered as a serious candidate for a description of elementary particle processes at the present stage of our understanding of the nature of fundamental interactions.

We shall present in Part III of our work the prediction of HFM for Large Hadron Collider experiments. We show there that predictions for various cross sections coming from Higgs-Free and Standard Models are drastically different so it will be easily to discriminate between these models.

ACKNOWLEDGMENTS

The authors are grateful to G. Altarelli, I. Białynicki Birula, B. Grządkowski, Z. Haba, M. Kalinowski, J. Werle and G. Veneziano for interesting discussions and S.D. Odintsov for sending to us the results of his group. They are especially grateful to S. Dittmaier for sending them his computer code and to D. Schildknecht for the extensive discussion of properties of his model.

## Appendix.

We prove here the Theorem 4.1. The measures in (3.17) have the form:

$$D\Phi = \prod_x d\Phi(x)$$

$$DA = \prod_{x,a,b,\mu,\nu,\sigma} dA_\mu^a(x) dW_\nu^b dB_\sigma(x) \qquad (A.1)$$

$$D\psi = \prod_{x,i} d\bar{\psi}_i(x) d\psi_i(x)$$

and according to [17]

$$Dg = \prod_{x,\mu \geq \nu} (-g(x))^{5/2} dg^{\mu\nu}(x) \qquad (A.2)$$

Let $\delta[g(\Psi)]$ be an another scale fixing condition. We show that the integral

$$Z^{-1} \int C(\Psi) e^{iS_T(\Psi)} \Delta_g(\Psi) \delta[g(\Psi)] D\Psi \qquad (A.3)$$

coincides with (3.15).

Note first that the measure $Dg$ is conformal invariant but the full measure $D\Psi$ given by (3.17) is not conformal invariant. It is crucial however that $D\Psi$ is multiplicatively conformal covariant. In fact from (2.3) it follows that

$$D\Psi^\Omega = \rho(\Omega) D\Psi \qquad (A.4)$$

where $\Psi_i^\Omega = \Omega^{s_{\Psi_i}} \Psi_i$ is the conformal transform of $\Psi_i$, $\rho(\Omega) = \prod_x \Omega^{N_\Psi}(x)$ and $N_\Psi = \sum_i s_{\Psi_i}$ is the sum of conformal degrees of scalar, fermion and vector fields.

Then from (3.15), (4,17) and (A.4) we have

$$< C(\Psi) >_0 = Z^{-1} \int C(\Psi) e^{iS_T(\Psi)} \Delta_f(\Psi) \delta[f(\Psi^\Omega)] D\Psi D\Omega$$

$$= Z^{-1} \int C(\Psi) e^{iS_T(\Psi)} \Delta_f(\Psi) \delta[f(\Psi)] \Delta_g(\Psi) \delta[g(\Psi^{\Omega'})] D\Psi \rho^{-1}(\Omega) D\Omega D\Omega' \quad (A.5)$$

Setting $\Psi^{\Omega'} = \Psi'$ and using the multiplicative covariance of $D\Psi$ measure and the invariance of $D\Omega$ measure we obtain

$$\int \rho^{-1}(\Omega) \rho^{-1}(\Omega') D\Omega = \int \rho^{-1}(\Omega \Omega') D\Omega = c \quad (A.6)$$

The same constant appears in the partition function $Z$ and these constants cancel out in (A.5). Hence using the invariance of $D\Omega$ under inversion $\Omega \to \Omega^{-1}$ and (3.19) we obtain

$$< C(\Psi) >_0 = Z^{-1} \int C(\Psi) e^{iS_T(\Psi)} \Delta_g(\Psi) \delta[g(\Psi)] D\Psi. \quad (A.7)$$